\documentclass[useAMS,usenatbib,usegraphicx]{mn2e}

\include{epsf}

\newcommand\aj{{AJ}}%
\newcommand\apj{{ApJ}}%
\newcommand\apjl{{ApJ}}%
\newcommand\apjs{{ApJS}}%
\newcommand\aap{{A\&A}}%
\newcommand\aaps{{A\&AS}}%
\newcommand\mnras{{MNRAS}}%
\newcommand\nat{{Nature}}%
\title[Metallicity gradients of disc stars for a cosmologically simulated galaxy]
{Metallicity gradients of disc stars for a cosmologically simulated galaxy}
\author[Rahimi et~al.]
 {Awat Rahimi,$^{1}$\thanks{E-mail: ara2@mssl.ucl.ac.uk}
Daisuke Kawata,$^{1}$
Carlos Allende Prieto,$^{1,2,3}$
Chris B. Brook,$^{4}$
\newauthor
Brad K. Gibson$^{4}$
 and 
Alina Kiessling$^{5}$
\\
$^{1}$ Mullard Space Science Laboratory, University College London,
Holmbury St. Mary, Dorking, Surrey, RH5 6NT
\\
$^{2}$Instituto de Astrof{\'i}sica de Canarias, 38205, La Laguna, Tenerife, Spain
\\
$^{3}$Departmento de Astrof\'{\i}sica, Universidad de La Laguna, 38206, La Laguna, Tenerife, Spain
\\
$^{4}$Jeremiah Horrocks Institute for Astrophysics and Supercomputing, University of Central Lancashire, 
Preston, PR1~2HE
\\
$^{5}$University of Edinburgh, Royal Observatory, Blackford Hill, 
Edinburgh, EH9 3HJ, UK
}
\date{Accepted .
      Received ;
      in original form }

\pagerange{\pageref{firstpage}--\pageref{lastpage}}
\pubyear{2011}

\begin{document}

\maketitle

\label{firstpage}

\begin{abstract}
We analyse for the first time the radial abundance gradients of the disc stars of a disc galaxy simulated with our three dimensional, fully cosmological chemodynamical galaxy evolution code {\tt GCD+}. We study how [Fe/H], [N/O], [O/Fe], [Mg/Fe] and [Si/Fe] vary with galactocentric radius. For the young stars of the disc, we found a negative slope for [Fe/H] and [N/O] but a positive [O/Fe], [Mg/Fe] and [Si/Fe] slope with radius. By analysing the star formation rate (SFR) at different radii, we found that the simulated disc contains a greater fraction of young stars in the outer regions, while the old stars tend to be concentrated in the inner parts of the disc. 
This can explain the positive [$\alpha$/Fe] gradient as well as the negative [N/O] gradient with radius.
This radial trend is a natural outcome of an inside-out formation of the disc, regardless of its size and can thus explain the recently observed positive [$\alpha/Fe$] gradients in the Milky Way disc open clusters.
\end{abstract}


\section{Introduction}
\label{intro-sec}
In recent years, there have been extensive efforts devoted to measuring chemical abundance trends within the Galactic disc \citep{fjt02, chw03, dc04, egp05, yct05, sbr08, mc09, pcr10, bao10} using various tracers. The aim of these studies has been to find out how our Galactic disc formed and evolved with time. Around the solar neighborhood, good progress has been made recently, including the derivation of age information for several clusters \citep[e.g.][]{pcr10} and individual stars \citep[e.g.][]{nma04, pe04, jl05, hna07, hna09}. However, further out, at larger galactocentric radii, there is a striking lack of any large high quality datasets.
To unravel the formation history of the Galactic disc, we need to know the abundance trends in the stars along the disc complemented with accurate age information.

In the near future ESA's Gaia mission should provide exceptional positional and proper motion information for up to one billion stars in the Milky Way. The vast majority (80\%) of these stars will lie in the disc. 
For bright stars, the Gaia RVS will determine the abundances of iron and the $\alpha$-elements. In addition, future and proposed ground-based projects including the Apache Point Observatory Galactic Evolution Experiment, APOGEE \citep{apms08} and the High Resolution Multi-object Echelle Spectrograph (HERMES) survey \citep{fb08} will provide more detailed and accurate chemical abundance determinations which will complement the Gaia data.

The chemical properties of stars are important to understand the formation history of the disc. Chemical elements heavier than boron are the end products of stellar evolution. The so-called $\alpha$-elements and iron (Fe) are of particular interest since it is known that they are produced primarily in Type II (SNe II) and Type Ia (SNe Ia) supernovae respectively. SNe Ia and SNe II have different timescales and thus studying the abundance ratios of the $\alpha$-elements with respect to Fe gives unique fossil information on the past conditions and evolution of the galaxy. 
In addition to studying various [$\alpha$/Fe] ratios, we also look at how the ratio of [N/O] varies with galactocentric radius as this can also be used as a cosmic clock. Good cosmic clocks are obtained by taking the ratios of elements produced by different stellar masses and therefore on different time-scales. Nitrogen is primarily produced by intermediate mass stars with longer lifetimes whereas oxygen is produced by SNe II progenitor stars which have a shorter lifetime. In any case, the actual abundance gradients and ratios of different elements will vary depending on the chemical evolution histories at different radii and therefore on the building-up history of the disc. 

The chemical evolution at different radii of the disc is studied theoretically using pure chemical evolution models \citep[e.g.][]{cmg97, cmr01, rkf05}. \citet{cmg97} suggested that the Milky Way forms primarily out of two infall episodes; the first giving rise to the halo and subsequently the bulge, and the second producing the disc via a much slower infall of primordial gas preferentially accumulating faster in the inner compared to the outer regions of the disc. This scenario is known as the ``inside-out" mechanism for disc formation. It is one of the possible mechanisms which can reproduce and explain the observed properties of the Milky Way. However in order to be able to draw firmer conclusions, it is necessary to have observational data for the outer regions of the Galactic disc \citep{cmr01}.

Inside-out formation may lead to some specific evolution of the abundance gradients. How do the abundance gradients evolve within the disc? There is still a lack of agreement between different authors; for some the gradients steepen with time \citep{cmg97}, whilst to others the gradients flatten with time \citep{hpb00, dc04, mc09}.
Note, however, that some authors have recently suggested that the traditional chemical evolution models such as those of \citet{cmr01} may need to be revised as they do not sufficiently consider the important effects of radial mixing \citep{sb09}.

In recent years, there has been progress in modelling disc galaxy formation in a Cold Dark Matter (CDM) Universe, using three dimensional numerical simulations \citep{nk92,sm94,bc00,anse03a,anse03b,bkgf04b,gwmb07,onb08,stw08,sws09}. Some studies include both SNe II and SNe Ia and discuss the details of the chemical properties in the simulated galaxies in an isolated halo collapse \citep{rvn96,b99} or hierarchical clustering \citep{bgmk05,rgm05,stw05,mss08} scenario. 
However, to date,  no three dimensional fully cosmological chemodynamical simulations have looked at radial abundance trends with age of disc stars. This is the first study to do so and it will be very interesting to see how our results will compare with the literature, both with theory and observations.

\begin{table*}
 \centering
 \begin{minipage}{140mm}
  \caption{Simulation parameters}
  \renewcommand{\footnoterule}{}
  \begin{tabular}{@{}lllllllllll@{}}
  \hline
   Name & $M_{\rm vir}$ & $r_{\rm vir}$ & $m_{\rm gas}$\footnote{Mass of gas per particle} & $m_{\rm DM}$ \footnote{Mass of DM per particle} & $e_{\rm gas}$ & $e_{\rm DM}$ & $\Omega_0$ & $h_0$ & $\Omega_{\rm b}$ \\
   & $(M_{\sun})$ & $(kpc)$ & $(M_{\sun})$ & {$(M_{\sun})$} & {$(kpc)$} & {$(kpc)$} & & & & \\
   \hline
Gal1 & $8.8\times10^{11}$ & 240 & $9.2\times10^{5}$ & $6.2\times10^{6}$ & 0.57 & 1.1 & 0.3 & 0.7 & 0.039  \\
 \hline
\end{tabular}
\end{minipage}
\end{table*}

 In this paper, we analyse the chemodynamical properties of the disc stars in a Milky Way size disc galaxy in our $\Lambda$CDM cosmological simulation.  Unfortunately, the simulated galaxy is not a late-type disc galaxy, such as the Milky Way. Nevertheless, since the detailed chemical distribution of disc stars is only available to us in the Milky Way, we compare our simulation with the Galaxy, and discuss what we could learn about the formation and evolution of a general disc component.

The outline for this paper is as follows. In Section 2 we describe our numerical simulation and define our disc stars. In Section 3 we present the results of our chemodynamical analyses. We initially study the properties and differences between accreted and {\it in situ} stars within the disc in Section 3.1. In Section 3.2 we analyse any radial abundance trends in the disc and its relation to the age of the disc stars with the aim of unravelling the formation mechanisms of our disc. Finally we present our conclusions in Section 4.

\section{The code and model}
\label{code-sec}

To simulate our galaxy, we use the original galactic chemodynamical evolution code {\tt GCD+} developed by \citet{kg03a}. 
{\tt GCD+} is a three-dimensional tree $N$-body/smoothed particle hydrodynamics code   \citep{ll77,gm77,bh86,hk89,kwh96} that incorporates self-gravity, hydrodynamics, radiative cooling, star formation, supernova feedback, and metal enrichment. {\tt GCD+} takes into account chemical enrichment by both SNe II \citep{ww95} and SNe Ia \citep{ibn99,ktn00} and mass loss 
from intermediate-mass stars \citep{vdhg97}, and follows the chemical enrichment history of both the stellar and gas components of the system. As described in \citet{kg03a} {\tt GCD+} takes into account the metallicity dependence of the age of stars \citep{ka97}, the metal dependent yields from SNe II and mass loss from intermediate mass stars.

Radiative cooling, which depends on the metallicity of the gas
 \citep[derived with {\tt MAPPINGSIII}:][]{sd93} is taken into account. The cooling rate for a gas with solar metallicity is larger than that for gas of primordial composition by more than an order of magnitude. Thus, cooling by metals should not be ignored in numerical simulations of galaxy formation \citep{kh98,kpj00}. However, we ignore the heating effect of the cosmic UV background radiation and UV radiation from hot stars for simplicity.

Star formation is modelled using a method similar to that suggested by \citet{nk92} and \citet{kwh96}. For star formation to occur, the following three criteria must be satisfied: (i) the gas density is greater than some critical density; (ii) the gas velocity field is convergent and (iii) the Jeans instability condition is satisfied. Our SFR formula corresponds to the Schmidt law. We assume that stars are distributed according to the \citet{s55} initial mass function (IMF). For more details see \citet{kg03a}.
Note that we assume only thermal energy feedback from SNe. It is known that the thermal energy feedback model has a negligible effect on galaxy formation \citep[e.g.][]{nk92,bkgf04a}. In real galaxies, the effects of SNe feedback are likely stronger, and could affect the chemical evolution in the disc component, as discussed in \citet{stw08}. However, there is still 
no model that can convincingly explain how SNe feedback affects galaxy formation and evolution. In this paper rather than exploring a variety of feedback modelling, we have implemented one simple feedback model and study how chemical evolution takes place under this assumption.
Our simulation, therefore, does not include any outflow, such as a Galactic fountain \citep{bre80}.
Although some authors \citep[e.g.][]{rmd01} suggest that SNe Ia dump more energy to the surrounding ISM than SNe II, we assume the same energy per supernova for SNe II and SNe Ia for simplicity.

\begin{figure}
\centering
\includegraphics[width=\hsize]{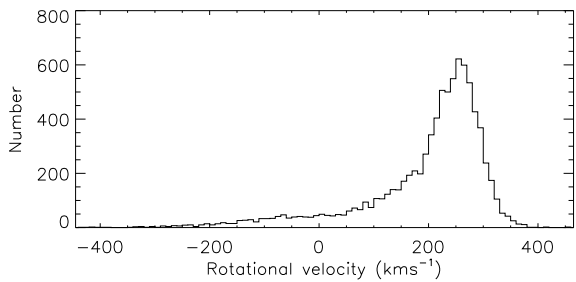}
\caption{
Histogram showing the rotational velocities of stars in our disc region. We defined disc stars to have rotational velocity between 150-350 kms$^{\rm-1}$.
The circular rotation speed in the central regions of our disc, at a radius of 7 kpc was 260 kms$^{\rm-1}$.
}
\label{vphihist-fig}
\end{figure}

The galaxy simulated here is from the sample of \citet{rah09}, referred to as ``Gal1". Gal1 is a high resolution version of galaxy ``D1" in \citet{kgw04}.  \citet{rah09} analyse a second galaxy (``Gal2"), but Gal2 has a lower resolution and a less prominent disc,  and thus we focus only on Gal1 in this paper. We used the multi-resolution technique in order to maximise the mass resolution within the regions where the disc progenitors form and evolve \citep{kgw04}. Here, only the high resolution region includes the gas particles, and therefore star formation.

We summarise the properties of our simulated galaxy in Table 1 adapted from \citet{rah09}. The second column represents the virial mass; the third column, the virial radius; Columns 4 and 5 represent the mass of each gas and DM particle in the highest resolution region, and Columns 6 and 7 are the softening lengths in that region.  The cosmological parameters for the simulation are presented in Columns 8--10. $\Omega_0$ is the total matter density fraction, $h_0$ is the Hubble constant (100 ${\rm kms^{-1}Mpc^{-1}}$) and $\Omega_{\rm b}$ is the baryon density fraction in the universe.
The age of the Universe is 13.5 Gyr in our simulation.

To identify the main progenitor galaxy, a friends-of-friends (FOF) group finder is used at regular time intervals in the simulation. Specifying a linking length $b$ and identifying all pairs of particles with a separation equal to or less than $b$ times the mean particle separation as friends, stellar groups are defined as sets of particles connected by one or more friendship relations. The other main parameter in the FOF algorithm is the minimum number of particles. By setting this parameter sufficiently high, one avoids including spurious objects that may arise by chance. In our simulations we use a linking parameter $b=0.01$ and a threshold number of particles of 100. We define the largest group which has the highest number of the FOF identified particles as the main progenitor galaxy.

\begin{figure}
\centering
\includegraphics[width=\hsize]{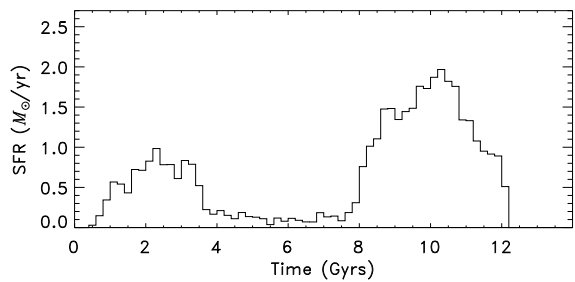}
\caption{
Star formation rate (SFR) history of disc stars . Two periods of star formation are visible. }
\label{totsfr-fig}
\end{figure}

Note that the version of {\tt GCD+} used in this paper \citep{kg03a} applies the SNe II yields calculated by \citet{ww95}. 
The iron yield is ambiguous for the SNe II nucleosynthesis model, as there is a large uncertainty in its value. It is well known that the iron yield shown in \citet{ww95} seems to lead to lower [$\alpha$/Fe] values, compared to those observed in low-metallicity stars in the solar neighborhood. Therefore, some authors commonly use half of the \citet{ww95} iron yield  \citep[e.g.][]{tww95,bg97,glm97}. The version of GCD+
 used in this paper applies the actual value of the iron yield in \citet{ww95}.
We, however, allow for this by only comparing the relative difference of [$\alpha$/Fe] among different samples of stars within our simulated galaxy.
Also note that
although we disperse metals to neighbour gas particles when a star dies, weighted by a kernel,
 metal diffusion between gas particles was not considered in this simulation.
Therefore, the spread in the metallicity distributions will be artificially high, while the peak of the distributions should be robust. 
Since we likely overestimate the scatter in our results, we take the median value for approximately every 100 stellar particles in each
[Fe/H], [O/Fe] or radius bin in Section~\ref{comp-acc-ins} and Section~\ref{age-analysis} respectively and plot it as a single point. This method reveals more clearly any trends present in the data.

We identify disc particles using the simulation output at the end of the simulation. We use the output at $z=0.1$ (t $\sim$ 12.3 Gyr), as going to any lower redshift results in an unacceptable amount of contamination from low-resolution particles in our simulated galaxy. 
To define the disc, we first set the disc plane of our galaxy to be along the gas axis (x-y plane) and the rotation axis to be the z-axis. The disc was defined as extending radially between 4 and 10 kpc in the galactic plane and $\pm1$ kpc in the z-direction. We did not include in our sample any stars found at less than 4 kpc from the centre of the galaxy to minimise possible contamination with bulge stars. Furthermore, only stars with a rotational velocity between 150-350 kms$^{\rm-1}$ were included in our disc sample. This velocity range was a somewhat arbitrary choice. These combined criteria, however, worked well to isolate disc stars.
In Fig.~\ref{vphihist-fig}, we show the histogram of rotation velocity for the stars in our disc region. Fig.~\ref{vphihist-fig} shows that the rotation component is clear and dominant in the disc region.
Nevertheless, our sample does include some bulge stars, because the simulated galaxy has a large fraction of bulge stars. However, the rotation velocity criterion restricts the contamination
only to the population of bulge stars which have a significant rotation velocity. Also note that in our simulated galaxy the bulge stars formed only at early epochs \citep{rah09}, so there should not be any contamination
for the younger stars that are the focus of this paper.

\begin{figure}
\centering
\includegraphics[angle=0,width=\hsize]{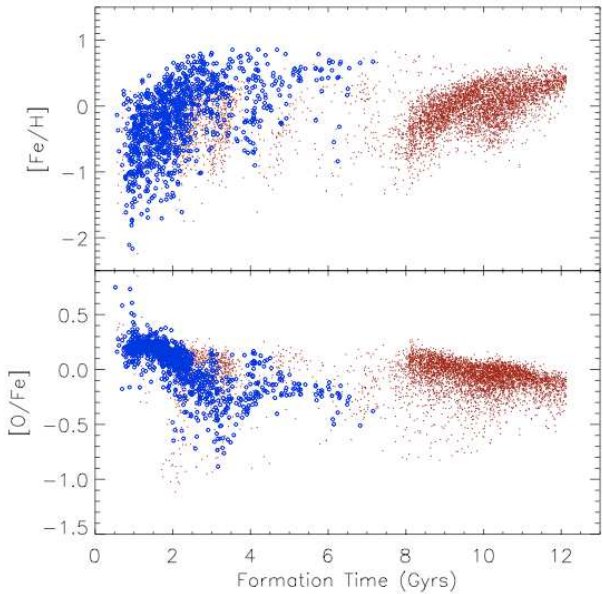}
\caption{
[Fe/H] (upper) and [O/Fe] (lower) against formation time ($t_{f}$) for our disc stars. Red filled and blue open circles are {\it in situ} and accreted disc stars respectively.
}
\label{feoh-fig} 
\end{figure}

\begin{figure}
\centering
\includegraphics[angle=0,width=\hsize]{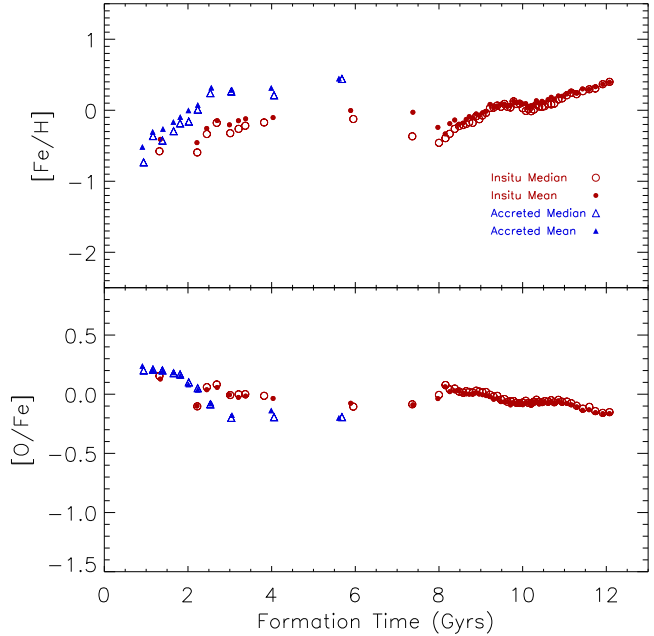}
\caption{
Same as in Fig.~\ref{feoh-fig}, however now plotting only the median and mean every 100 particles to make the trends in our data clearer. The open and filled circles represent the median and mean for {\it in situ}  stars respectively and the open and filled triangles the median and mean for accreted stars respectively.
}
\label{meanvsmed-fig} 
\end{figure}

\begin{figure}
\centering
\includegraphics[width=\hsize]{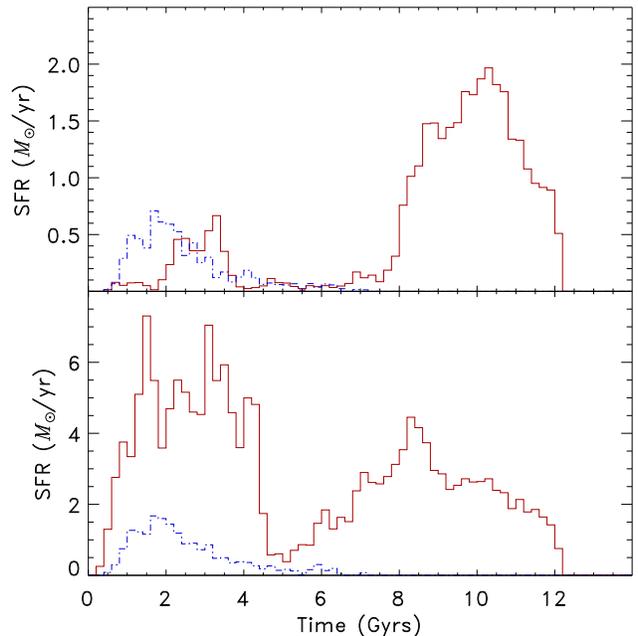}
\caption{
SFR history of our defined disc stars (upper panel) showing the contributions from {\it in situ} and accreted stars (red solid and blue dot-dashed lines respectively). 
In the lower panel we replot the SFR for our sample of disc stars but now in the region 0 $<$ $R_{\rm GC}$ $<$ 10 kpc  and $|z|$ $<$ 1 kpc, with a galactic rotation velocity between 150 and 350 kms$^{\rm-1}$.
}
\label{sfr-fig}
\end{figure}

\begin{figure*}
\centering
\includegraphics[angle=0,width=\hsize]{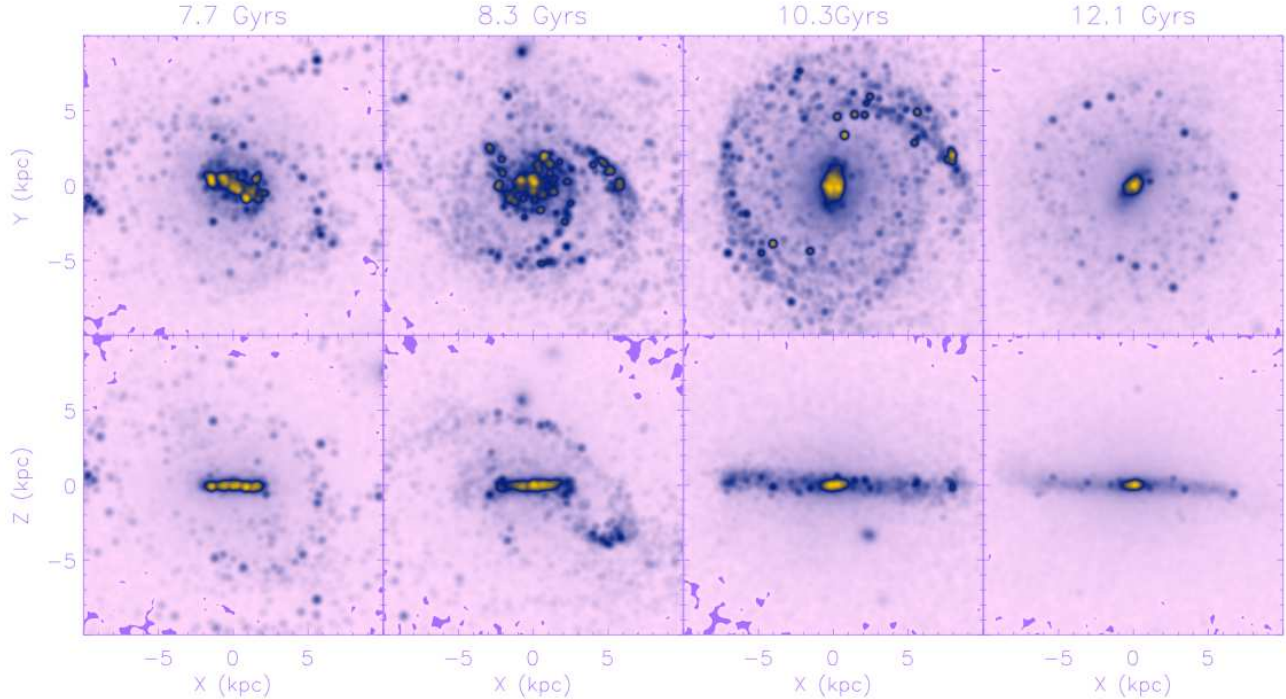}
\caption{
Face-on (upper) and edge-on (lower) view of the evolution of the galaxy from t = 7.7 - 12.1 Gyr, colour coded by the expected {\it V} band luminosity \citep{bc03}. The brightest regions are the densest. The knots are due to young bright particles.
}
\label{luminmap-fig}
\end{figure*}

\begin{figure*}
\centering
\includegraphics[angle=0,width=\hsize]{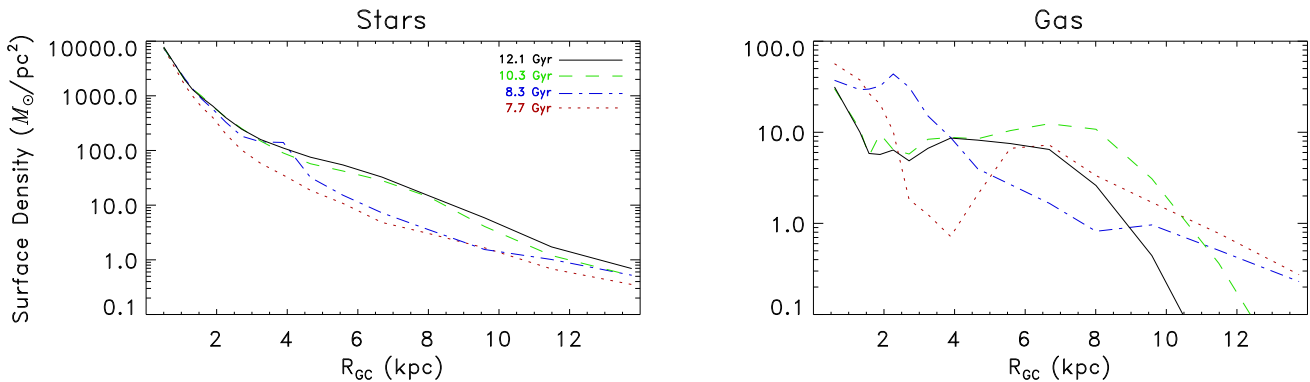}
\caption{
Surface density profile of stars (left) and gas (right) in the galaxy from t = 7.7 - 12.1 Gyr. 
}
\label{surfden-fig}
\end{figure*}

\section{Results}
\label{res-sec}

Fig.~\ref{totsfr-fig} shows the total SFR as a function of time for our sample of disc stars.
Note that Fig.~\ref{totsfr-fig} does not include particles which were born in the disc region, but at this particular time are not in the disc. We expect that a significant amount of stars, especially old stars, are kinematically heated up after they formed in the disc, due to our poor resolution. Therefore, we likely underestimate the population of disc stars, and our results focus on the population of relatively cold disc stars at the final time step.
    From Fig.~\ref{totsfr-fig} we see that our simulated galaxy had two episodes of star formation, the first occurring at an early epoch centred around 2 Gyr and ending before 4 Gyr. The other main episode of star formation occurred at much later times and was more intense and lasted longer.
The major episode of star formation responsible for making the disc started after 7 Gyr and lasted till the final timestep. 
Note that our simulated galaxy stops at $z = 0.1$. 
Also note that the overall star formation history is not like that inferred from solar neighbourhood stars in the Milky Way, although there is no consensus as to what that is exactly \citep[e.g.][]{bn01,rpf04,ab09}. 
This also shows that our simulated galaxy is different from the Milky Way. We bear this in mind in the following discussion.

Our simulated galaxy formed through hierarchical clustering, and early disc formation is associated with a series of mergers at early epochs when bulges are also built up \citep{rah09}. 
One of the aims of this study is to show how we can infer such a merger history and building up history of the disc from the present-time properties of disc stars. To this end, first we separate the disc stars into accreted stars and stars formed {\it in situ}. From the current chemical and kinematical properties of  disc stars, we study how
chemical and kinematical properties differ between accreted and {\it in situ} stars in Section~\ref{comp-acc-ins}. In Section~\ref{age-analysis} we split our {\it in situ} disc stars into two samples based on the age of the stars. We then compare the properties of our two groups of disc stars at different galactocentric radii to see if we can find any metallicity gradients with galactocentric radius and how they relate to the formation history of the discs. Finally, in Section~\ref{conc}, we summarise our findings.

\subsection{Accreted and {\it in situ} stars}
\label{comp-acc-ins}

In this section, we compare the chemical and kinematical properties of accreted and {\it in situ} stars. We trace back the formation time and location for all the disc stars, and any stars that are born within a radius of 20 kpc from the largest progenitor and end up within the area defined as the disc at the final timestep are given the title ``{\it in situ} stars". 
Stars that form at a radius greater than 20 kpc from the centre of the galaxy and end up within the disc at the final timestep, we term ``accreted stars". We chose 20 kpc arbitrarily because we discuss only rough trends between the two populations. We experimented using a larger radius up to 30 kpc, and generally found the same conclusions.


Fig.~\ref{feoh-fig} shows [Fe/H] and [O/Fe]  versus formation time for accreted and {\it in situ} stars. 
The evolution of disc metallicity with time has been observed in disc stars of the Milky Way \citep{rtl03, bfl04,bf06} although only with small number statistics.
Oxygen is one of the so-called $\alpha$-elements. These elements are primarily produced in massive stars with short lifetimes that explode as SNe II. Iron is produced predominantly in SNe Ia, from lower mass binary stars with longer lifetimes \citep[we apply the model proposed by][]{ktn00}. Accreted stars have lower [Fe/H] and higher [$\alpha$/Fe] since in this particular galaxy they form in early epochs before the enrichment from SNe Ia becomes important. {\it In situ} stars continue to be born up until the final time-step of the simulation.
Fig.~\ref{meanvsmed-fig} shows the same information but taking the median and mean every 100 particles to show the trends more clearly.
There is only a very small difference between the two methods (especially for the  younger stars which are the focus of this paper).

In the upper panel of Fig.~\ref{sfr-fig}, we replot the SFR history for our sample of disc stars now including the individual contributions from {\it in situ} and accreted stars. We see that the accreted stars contribute prominently to the fraction of old stars in our sample. At later times we see that the disc grows due to the formation of stars {\it in situ} in the disc. {\it In situ} stars therefore are mainly responsible for the formation of the disc at times later than 7 Gyr. This is closely related to the formation time of the bulge of this galaxy, which takes place over the first few Gyr \citep[see Fig. 1 of][]{rah09}. We can therefore deduce that the disc of our simulated galaxy only really begins to build up after the bulge is fully formed. However, Fig.~\ref{sfr-fig} also demonstrates that some disc stars can form {\it in situ} during the formation epoch of the bulge stars due to mergers. It is possible that some of these stars may have formed in the vicinity of the bulge, but ended up in the disc at the final timestep \citep{rdq08,scg09}.
In the lower panel of Fig.~\ref{sfr-fig}, we replot the SFR history as in the upper panel, but now including the central 4 kpc of the bulge component. We see that star formation in the inner region starts
 earlier than in the outer region.
In Fig.~\ref{luminmap-fig} we show how our simulated galaxy evolves from t = 7.7 - 12.1 Gyr. We can clearly see that before 8 Gyr the radial extent of the galaxy is less than 4 kpc. Only after 8 Gyr does the disc grow beyond 4 kpc. This explains why in Fig.~\ref{totsfr-fig} and the upper panel of  Fig.~\ref{sfr-fig}, it appears as if the galaxy suddenly starts forming stars at around t = 8 Gyr. It is simply due to our selection criteria for the extent of the disc region.
Fig.~\ref{surfden-fig} shows the surface density radial profiles of both the stars and gas for our galaxy corresponding to the times given in Fig.~\ref{luminmap-fig}. 
The stellar surface density of the inner region ($R_{\rm GC}$ $<$ 4 kpc) increases earlier (t $<$ 8.3 Gyr) than the outer region.
From these figures, we deduce that the simulated galaxy formed inside-out. In our defined disc region, the gas density is greatest at 10.3 Gyrs. This corresponds to the peak of the star formation rate in the disc.

\begin{figure}
\centering
\includegraphics[angle=0,width=\hsize]{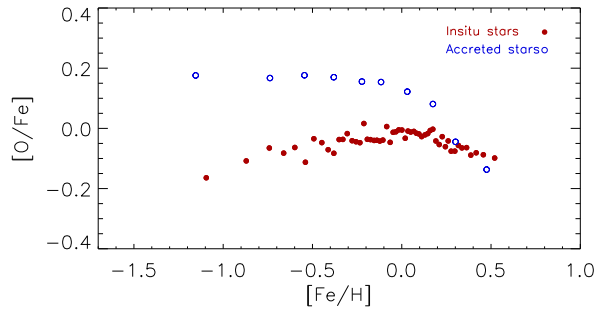}
\caption{
[O/Fe] versus [Fe/H] at the final timestep. Red filled and blue open circles are {\it in situ} and accreted stars respectively. As expected from Fig.~\ref{feoh-fig} accreted stars have higher [O/Fe] ratios. 
}
\label{ofevsfeh-fig}
\end{figure}

In Fig.~\ref{ofevsfeh-fig} we plot the median of [O/Fe] versus [Fe/H], using the median value for every 100 particles as described in Section~\ref{code-sec}. Here, the abundance of oxygen is used to represent $\alpha$-elements. Accreted stars have higher [$\alpha$/Fe] and lower [Fe/H] compared to {\it in situ} stars, corresponding to them being formed earlier. 
We find that 90\% of the accreted stars fell into the region within a radius of 10 kpc before $t=8$ Gyr. This is before the disc starts forming at radii greater than 4 kpc  (Figs.~\ref{feoh-fig} and~\ref{sfr-fig}).
As a result there exists a distinct population of accreted stars with high [$\alpha$/Fe]. These accreted stars likely end up in a thick disc \citep{anse03b,hnnh06}.
In fact, the velocity dispersion of the vertical component of the disc is 96 and 58 kms$^{\rm-1}$ for accreted and {\it in situ} stars respectively in the simulated galaxy.
Note that this is significantly higher than what is observed around the solar neighbourhood \citep[e.g.][]{hna09}. 

\begin{table}
 \centering
\begin{minipage}{70mm}
\caption{Abundance gradients for various element abundance ratios for the intermediate and young stars}
\begin{tabular}{lll}
\hline
\multicolumn{3}{c}{Abundance Trends (dex ${\rm kpc}^{-1}$)} \\
\hline
Abundance ratio & Intermediate & Young \\ \hline
 [Fe/H] & $-0.050$ & $-0.066$ \\ \hline
[O/Fe] & $-0.005$ & 0.009 \\ \hline
[Mg/Fe] & $-0.004$ & 0.012 \\ \hline
[Si/Fe] & $-0.004$ & 0.006 \\ \hline
[N/O] & $-0.005$ & $-0.012$ \\ \hline
\end{tabular}
\end{minipage}
\end{table}

 Our chemodynamical simulation demonstrates that if such accreted stars are formed {\it exclusively} at early epochs, then an accretion origin for the thick disc can explain the observed distinct difference in  [$\alpha$/Fe] between the thick and thin disc in the Milky Way \citep[e.g.][]{gcm96, gcm00, fuh98, pnc00, tet01, fbl03, rtl03, sp03, bfl05, fuh08}.
 We note however that it is more difficult to explain the thick disc stars with higher metallicity as the accreted stars, because the accreted stars formed in smaller galaxies that are likely to be more metal poor. Although our simulation shows high metallicity and low [O/Fe] accreted stars, since we assume a negligible SNe feedback model (Sec. 2) and ignore the UV background
radiation, our simulation likely overestimates the metallicity of the accreted stars formed in such small galaxies \citep[e.g.][]{gwmb07,bgb07}. The majority of thick disc stars in the Milky Way are relatively metal rich, and 
may require more intense star formation which may be more readily associated with the {\it in situ}  population \citep[e.g.][]{cmr01, bkgf04b,sbin09}.

\begin{figure}
\centering
\includegraphics[width=\hsize]{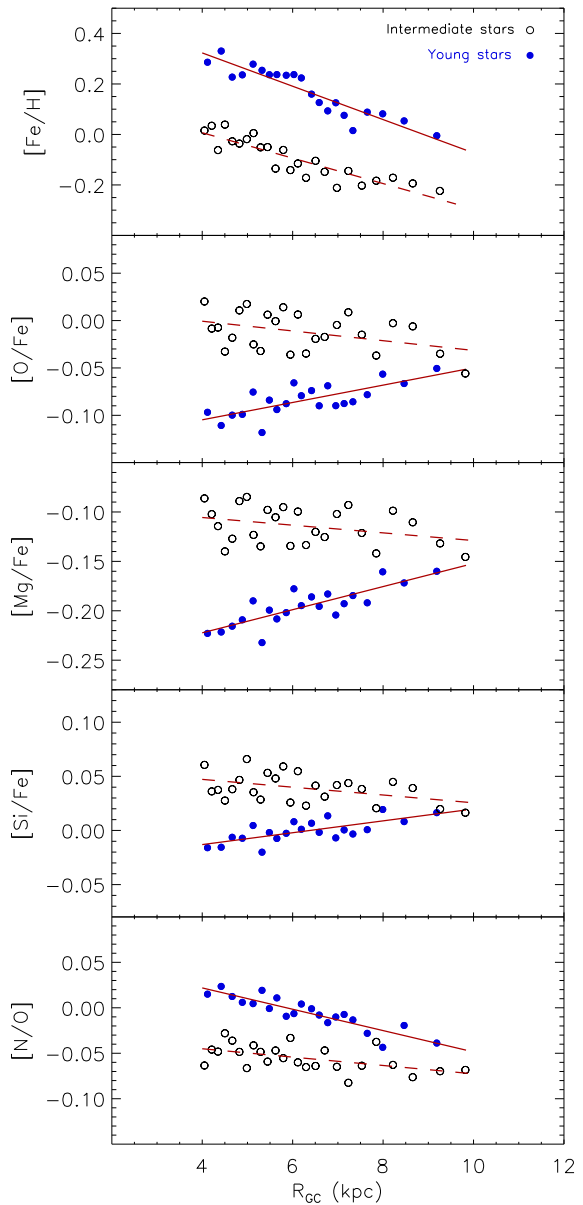}
\caption{
[Fe/H], [O/Fe], [Mg/Fe], [Si/Fe] and [N/O] versus galactocentric radius. The black open and blue filled circles represent intermediate-age and young stars respectively. The red dashed and solid lines are the fitted linear function to the median of the intermediate and young stars respectively whose slopes are listed in Table 2.
}
\label{abund-fig}
\end{figure}

\subsection{Radial trend of chemical properties and age dependence}
\label{age-analysis}

In our simulation, we follow the formation history of our galaxy from very early times to the final timestep of the simulation ($z=0.1$). Since the differently aged populations in the galaxy likely hold different memories of the formation history, we analyse the properties of stars with different ages.
Fig.~\ref{totsfr-fig} clearly shows that our simulated galaxy has two periods of disc star formation. The second and major episode is of particular interest as this is what predominately leads to the formation of disc stars from smooth gas accretion. Prior to this epoch, multiple mergers prevent smooth gas accretion to form a disc \citep{rah09}. 
The second and major period
lasts from t = 7 to 12 Gyr and is characterised by a rapid rise and fall in the SFR.
It would be interesting to consider the differences between the stars formed at the two different parts of this period of star formation. Also, since we later would like to compare our results with the relatively young stars found in the Galactic disc, we therefore decided to divide
 the stars formed {\it in situ} during the second episode into 
 ``intermediate" disc stars (7 $<$ $t_{f}$ $<$ 10 Gyr) and ``young" disc stars formed during the final 2 Gyr ($t_{f}$ $>$ 10 Gyr). 
Note that in this section, we only focus on {\it in situ} stars, to track the building up history of the disc.

\begin{figure*}
\centering
\includegraphics[width=\hsize]{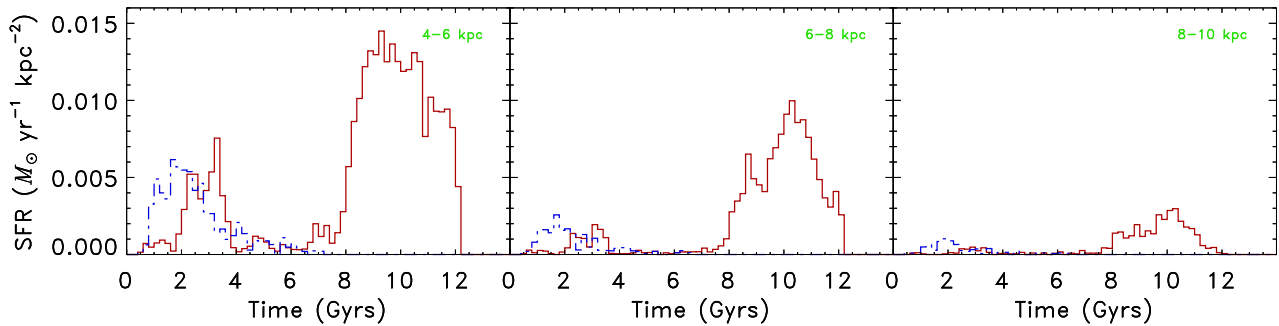}
\caption{
SFR versus time at different radial regions along the disc plane
\vspace{2cm}}
\label{sfr1-fig}
\end{figure*}

In the top panel of Fig.~\ref{abund-fig} we plot [Fe/H] against radius from the galactic centre, $R_{\rm GC}$, for intermediate-age and young stars. Here again we have plotted the median for every 100 particles in each radial bin. The straight and dashed red lines represent the best fit to the data. 
Here we only consider the disc stars within 10 kpc, because in the outer region there are too few particles to represent the lower density regions of the disc.
We clearly see a negative [Fe/H] slope with $R_{\rm GC}$. 
 The younger the stars, the higher the median metallicity at any given $R_{\rm GC}$. Table 2 shows the fitted slope to the median abundances for the intermediate and young stars.
 Note that the exact value of these gradients are not important and depend slightly on how we measure or sample the data. Below we only discuss qualitative trends of these gradients.
  The intermediate and young stars have a similar slope in [Fe/H], although for the young stars the slope is slightly steeper.
Therefore we find that during the major disc formation phase there is little evolution of the metallicity gradient, although the metallicity increases with time. This seems to be consistent with what was found by model A of \citet{cmr01} in the latter epoch of evolution. However note that our cosmological simulations involve more complex processes, such as radial mixing of gas and stars, and a complex gas accretion history.
 For the young stars of the disc we clearly see a negative slope in [Fe/H] with $R_{\rm GC}$. This kind of detailed abundance observations along the disc is only really available for our Milky Way. Interestingly, various observational studies \citep{taa97, fjt02, yct05, sbr08, msr09, pcr10} have found similar abundance trends for young stars to what we see in our simulated galaxy.

\begin{figure}
\centering
\includegraphics[width=\hsize]{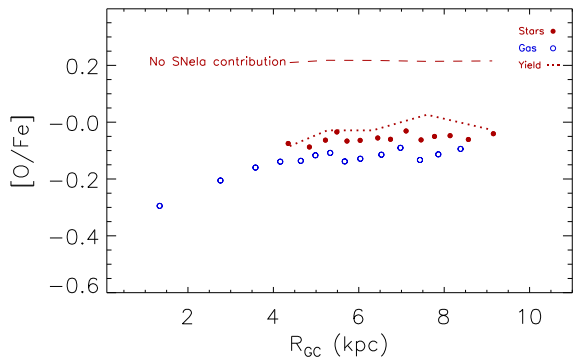}
\caption{
[O/Fe] versus galactocentric radius at 11.2 Gyr. The blue open circles and red dots represent gas and young stars respectively. The small red dots represent the yield from the disc stars in the last 1 Gyr. The red dashed line redraws this yield excluding the contribution from SNe Ia.  
}
\label{gasnstarofe-fig}
\end{figure}

The second row of Fig.~\ref{abund-fig} shows [O/Fe] against $R_{\rm GC}$ using the same plotting conventions and Table 2 shows the slope in the median abundances for both the intermediate and young stars. Generally speaking we see a slope which is close to being consistent with flat. However for the young stars, we do see a slight positive trend in the median.
This positive slope is also observed for Galactic open clusters in \citet{yct05} and \citet{pcr10}. 
One should note that the age range for the young stars roughly corresponds to the \citet{yct05} and \citet{pcr10} open clusters.
Also note however that the \citet{yct05} sample had a stronger positive gradient than the \citet{pcr10} sample, and that the actual slope is still uncertain.
\citet{cmr01} showed that infalling gas from the halo can impact the metallicities in the outer regions of the disc.
In Fig.~\ref{gasnstarofe-fig}, we analyse [O/Fe] for gas and stars younger than 1 Gyr at t = 11.2 Gyr when young stars in Fig.~\ref{abund-fig} are forming. Fig.~\ref{gasnstarofe-fig} shows that stars inherit their abundance patterns from the gas. We also plot [O/Fe] for the yield in the last 1 Gyr from our disc stars, i.e. stars at $|z|$ $<$ 1 kpc and 4 $<$ $R_{\rm GC}$ $<$ 10 kpc  and rotating with 150 $<$ $V_{rot}$ $<$ 350 kms$^{\rm-1}$. 
More than 70\% of stars in the disc region, i.e. 4 $<$ $R_{\rm GC}$ $<$ 10 kpc and $|z|$ $<$ 1 kpc satisfy our criteria for the rotation velocity of the disc stars, yet 93\% of the mass fraction of metals is ejected from such disc stars, due to the young age of this population. It is therefore apparent that the yields from our disc stars are the dominant source of chemical enrichment, and the yield from the spherical component is negligible in our
simulation\footnote{In Fig. 11, [O/Fe] of stars, gas and yields are slightly different
from each other. However, stellar abundances come from young stars which formed in
the last 1 Gyr, yields are only for the last 1 Gyr, and the gas is the
abundance at 11.2 Gyr. In addition, the plotted stellar and gas metallicities 
are median values at each radial region. Stars formed only
from high-density gas. Considering these facts, it is not surprising
that they are different.}.

Abundance ratios, such as [O/Fe] are very sensitive to the star formation histories.
In order to investigate how the star formation history changes with $R_{\rm GC}$, and thereby better understand these trends, in Fig.~\ref{sfr1-fig} we plotted the SFR against time for three different $R_{\rm GC}$ bins along the disc. Our three regions, going radially outwards were defined as extending from $4-6$ kpc, $6-8$ kpc and $8-10$ kpc. The magnitude of the SFR decreases with increasing $R_{\rm GC}$.  By 8-10 kpc (3rd panel from the left) there is a significant reduction in the magnitude of the SFR. More interestingly, however, the peak of the SFR occurs at slightly later times as $R_{\rm GC}$ increases.
Together with Figs.~\ref{sfr-fig},~\ref{luminmap-fig} and~\ref{surfden-fig} this provides
clear evidence for the inside-out formation of the disc. 
As a result, there is a higher fraction of young stars in the outer regions of the disc. This can explain the [O/Fe] enhancement because a more significant enrichment from SNe II is occurring in the outer region. On the other hand, in the inner region, there are relatively more old stars which are the precursors of SNe Ia (producing more Fe). 
Fig.~\ref{gasnstarofe-fig} also shows the yield coming exclusively from SNe II, which show a
constantly high [O/Fe] at all radii. This also indicates that the difference in the significance of SNe Ia enrichment is driving the positive [O/Fe] gradient.
Therefore we conclude that inside-out formation of the disc naturally leads to a higher fraction of young stars in the outer regions and a positive [$\alpha$/Fe] slope in the young population of the disc.

In Fig.~\ref{abund-fig} we also show how [Mg/Fe] and [Si/Fe] vary with $R_{\rm GC}$ and Table 2 shows the value for the median fitted slopes for the intermediate and young stars.
Magnesium and silicon are also $\alpha$-elements, and are mainly produced in massive stars. Therefore we observe similar trends to the [O/Fe] case. However, for the young stars, [Si/Fe] shows a slightly flatter slope compared to [O/Fe] and [Mg/Fe] cases. This could be because silicon is also significantly produced by SNe Ia \citep[we adopt the yields in][]{ibn99} compared to SNe II as also shown in \citet{glm97}. 

We also analyse [N/O] as shown in the final row of Fig.~\ref{abund-fig} and the fitted slopes are presented again in Table 2. [N/O] is expected to show an opposite trend to the [$\alpha$/Fe] case. Accordingly, for the young stars we see a negative slope that has an opposite sign to the [$\alpha$/Fe] case. From Fig.~\ref{sfr1-fig}, this is because at the outer radii, a greater fraction of young stars exist which end up as SNe II and thereby produce more oxygen making [N/O] lower. On the other hand, more intermediate mass stars \citep[we adopt yields in][]{vdhg97} died in the inner regions, producing more nitrogen, which makes [N/O] higher. 

As mentioned above, in the Milky Way, [$\alpha$/Fe] in open clusters also shows a tentative trend of increasing with radius \citep{yct05, pcr10}. If the chemical composition of open clusters represents the properties of the field disc stars at similar galactocentric radii, our simulation demonstrates that the observed abundance trends can be explained by a higher fraction of young stars in the outer region due to an inside-out formation of the Galactic disc. 
Note that although we rely on the yields from SNe II, SNe Ia and intermediate
 mass stars as explained in Section 2, some of these values are still controversial
 and there are a variety of yields suggested by several groups. Some different
 yield sets may be able to lead to the positive [O/Fe] gradient
 without having inside-out star formation. In addition, if somehow
 the infalling gas strongly contributes to the chemical abundance in the
 gas disc, and has lower [O/Fe] in the inner region, this can also explain
 the positive [O/Fe] gradient. Needless to say, there are more scenarios
 that can explain the positive [O/Fe] gradient. Our scenario is not a unique solution,
 but our simulation demonstrates one possible explanation (derived with a fully cosmological simulation) which is also
 naturally predicted from disc galaxy formation in a $\Lambda$CDM universe.



\section{Summary} 
\label{conc}

In this study we have analysed the chemistry and the dynamics of the disc stars in a cosmologically simulated disc galaxy. The galaxy was similar in mass and size to the Milky Way, and contained distinct gas and stellar disc components \citep{bkg05,ckbtg06}.
 
The simulated galaxy showed two episodes of star formation which led to the buildup of the stellar disc. The first occurred at very early epochs and was mainly due to accreted stars being brought into the galaxy as a result of mergers. These early mergers also built up the bulge  \citep{rah09}. Therefore, our simulation demonstrates that some stars accreted earlier can become disc stars, if they happen to have the right angular momentum. Our simulation also has old stars formed {\it in situ}. It is interesting to note that even during the mergers that are mainly building up the bulge some disc stars can form {\it in situ} and stay in the disc for a long time. Intriguingly, in this galaxy, a fraction of the bulge stars have a significant rotational velocity component \citep{rah09}. 

The second episode of star formation starts after the mergers cease and continues till the final timestep. In this period, smooth gas accretion builds up the disc and stars formed {\it in situ}.
We have analysed the radial trend of [Fe/H], [O/Fe], [Mg/Fe], [Si/Fe] and [N/O] for these {\it in situ} stars, especially focussing on the relatively young stars ($t_{f}$ $<$ 2 Gyr). Our simulated galaxy shows a negative [Fe/H] gradient with $R_{\rm GC}$. Interestingly, the Milky Way has a similar trend as recently observed for open clusters \citep{fjt02, pcr10} and cepheids \citep{lga03, alm04, lfp08, pbs09}.
More interestingly, we also found a positive [$\alpha$/Fe] gradient for the young stars in the disc. This trend is similar to the recently observed [$\alpha$/Fe] gradient for open clusters in the Milky Way \citep{yct05, pcr10} as well as to predictions from some chemical evolution models \citep[e.g.][]{msr09}.

We find that inside-out formation of the disc can naturally produce such negative [Fe/H] and positive [$\alpha$/Fe] gradients.
In our simulated galaxy, the magnitude of the SFR declines in the outer disc.
Thus we induce a more progressed enrichment in the inner region and negative [Fe/H] slope.
We also find that the peak of the SFR occurs at a later epoch in the outer regions of our simulated galaxy, which is clear evidence for inside-out disc formation.
As a result, the outer region in the disc harbours a greater fraction of young stars that produce more SNe II and cause higher [$\alpha$/Fe].
This also leads to a negative slope of [N/O], because in the inner region a greater fraction of low mass stars produce more nitrogen. 
These results demonstrate that such radial gradients of chemical abundances are sensitive to the formation history of the disc.

Note that our simulated galaxy is not a late-type galaxy like the Milky Way, but is more like an early type disc galaxy. Therefore, the age distribution of the disc stars in our simulated galaxy is very different from the Milky Way disc stars. However we should be able to apply our simple conclusion to any disc that formed {\it in situ} as a result of smooth gas accretion. 
Therefore, we suggest that the observed positive [$\alpha$/Fe] gradients in the Milky Way disc stars can be explained if the Milky Way also experienced a clear inside-out formation and harbours a greater fraction of young stars in the outer region.
Although this is not a unique scenario and we do not reject other scenarios, this is naturally expected in a $\Lambda$CDM universe.

Recently, \citet{wdd09} found clear evidence for inside-out disc formation in M33 and claimed that there is a greater fraction of old stars in the inner regions of the disc of M33. Various authors have found that the metallicity gradient in the M33 disc decreases going away from the centre \citep[e.g.][]{mcg07,mvm07,vdm07,rs08,rsc08,msv09,cio09}. More measurements of abundance ratios across the M33 disc would be extremely interesting \citep[such as: e.g.][]{ebp09}. Recently, there have also been attempts to measure the chemical properties along the disc radius for disc  galaxies other than our Milky Way \citep[e.g.][]{rfg05, mgc09}.
Interestingly, \citet{yd08} find a tentative detection of more $\alpha$-enhanced populations and younger luminosity weighted mean ages in the outer disc region of the disc galaxy FGC 1440.

Our results are useful for comparing observations to our simulation where we can trace the formation history. However, we also admit that the current chemodynamical simulation model should be improved. For example, since we do not allow metal mixing between particles, we likely overestimate the scatter of the metallicity distribution and [$\alpha$/Fe] at different radii. We are now working on improving our chemodynamical model, and new simulations will provide valuable information to disentangle the formation history of the disc galaxies from current and future observations.

\section*{Acknowledgments}

AR, BKG, CBB and DK acknowledge the support of the UK's Science \& Technology
Facilities Council (STFC Grant ST/H00260X/1, ST/F002432/1 ). 
BKG, CBB and DK acknowledge the support of the Commonwealth Cosmology Initiative.
AK acknowledges the support of the European DUEL RTN, project 
MRTN-CT-2006-036133.
We acknowledge CfCA/NAOJ and JSS/JAXA where the numerical computations for this paper were performed. We thank Kate Pilkington and Francesco Calura for helpful suggestions.
Finally, we thank the anonymous referee for providing us with useful comments and suggestions.


\label{lastpage}

\end{document}